\documentstyle[editedvolume,epsf]{crckapb}

\newcommand{\etal}{\mbox{et~al.}}
\newcommand{\kms}{\mbox{\,km\,s$^{-1}$}}
\newcommand{\Mpc}{\mbox{\,h$^{-1}$\,Mpc}}
\newcommand{\gs}{\mathrel{\raise0.3ex\hbox{$\scriptstyle >$}\kern-0.70em
                 \lower0.71ex\hbox{{$\scriptstyle \sim$}}}}
\newcommand{\ls}{\mathrel{\raise0.3ex\hbox{$\scriptstyle <$}\kern-0.70em
                 \lower0.71ex\hbox{{$\scriptstyle \sim$}}}}

\begin{opening}
\title{WIDE FIELD SPECTROSCOPY AND THE UNIVERSE}
\author{MATTHEW COLLESS}
\institute{Mount Stromlo \& Siding Spring Observatories\\
           The Australian National University\\
	   Canberra, ACT, Australia}
\end{opening}

\runningtitle {WIDE FIELD SPECTROSCOPY AND THE UNIVERSE}

\begin{document}

\section{Introduction}

In this short survey of the applications of wide-field, multi-object
spectroscopy to galaxy evolution, large-scale structure and cosmology,
I interleave summaries of the general goals and state of play in these
fields with specific examples based on my own recent work. I first
briefly review the goals and figures of merit for current and future
redshift surveys, before examining some recent results from deep
surveys in the field, cluster redshift surveys and surveys of large
scale structure. I take a look beyond redshift surveys to other probes
of galaxy evolution before concluding with a discussion of the future
prospects for such studies using the next generation of wide-field,
multi-object spectrographs on large telescopes.

The main goals of extragalactic/cosmological studies for which
wide-field, multi-object spectroscopy is relevant are (i)~the world
model: $H_0$, $q_0$, $\Omega$, $\Lambda$, etc.; (ii)~large-scale
structure: $P(k)$, $\xi(r)$, Gaussianity, topology, morphology,
biassing, etc.; and (iii)~galaxy formation and evolution: luminosity
function and $\xi(r)$ as functions of galaxy type, environment and
lookback time, the redshift of peak star formation in bulges and
disks, etc.  Redshift surveys are essential to some of these topics
and one powerful alternative in studying others, either because they
are intrinsically population studies or because they demand large
statistical samples. In either case one is playing a numbers game, so
that the most efficient observational approach will employ wide-field
multi-object spectroscopy.

\section{Current Results from Redshift Surveys}

There has been much recent progress in studies of galaxy formation and
evolution. The general form of the evolution of the galaxy population
out to $z$$\sim$1 has begun to emerge from deep redshift surveys of
$\sim$10$^3$ field galaxies to $B$=24, $I$=22 and $K$=20 (Ellis \etal\
1996, Lilly \etal\ 1995, Cowie \etal\ 1996). These studies have
revealed differential evolution of the galaxy luminosity function with
luminosity and also with colour or spectral type. At the same time,
significant advances have been made in detecting high-redshift
($z$$\gs$3) star-forming galaxies directly from their stellar light
and not from an associated QSO or radio source (Steidel \etal\ 1996).

At the other end of the scale, several redshift surveys of
$\sim$10$^4$ nearby ($z$$\ls$0.1) galaxies have been carried out over
large volumes of the local universe in order to map large-scale
structure (CfA, SSRS, IRAS PSCz, LCRS, ORS, etc.). These surveys have
measured the galaxy distribution's power spectrum, $P(k)$, or
correlation function, $\xi(r)$, with fair precision on scales less
than about 100~\Mpc, but also reveal still larger structures,
suggesting that we have not yet mapped a sufficiently large volume to
constitute a fair sample of the universe.
 
Studies measuring both the redshift and the distance for $\sim$10$^3$
galaxies at redshifts $cz$$\ls$5000\kms\ have mapped the peculiar
motions in the local universe and thereby constrained the local mass
distribution and the nature of the missing mass. Such observations
are much more difficult than simple redshift surveys, as the usual
distance indicators are based on versions of the linewidth--luminosity
relation (the Fundamental Plane for ellipticals, the Tully-Fisher
relation for spirals) and so require measurements of the internal
motions of each galaxy. As discussed below, such studies are beginning
to be pushed to higher redshifts in order to use the
linewidth--luminosity relations as probes of galaxy evolution.

The use of galaxy redshift surveys to probe the parameters of the
world model, popular twenty years ago, has largely been abandoned
today due to the realisation that evolutionary effects on the galaxy
population swamp the more subtle effects of spacetime geometry.

There are at least two other important types of redshift survey which
I do not touch on at all here, namely surveys of QSOs (to probe their
evolution and {\em very} large-scale structures) and surveys of QSO
absorption line systems (ALS) and their associated galaxies (which
offer an alternative approach to studying galaxy evolution at high
redshift).

\section{Figures of Merit for Surveys}

Table~1 gives a brief summary of some recent and proposed redshift
surveys, along with the diameter of the telescope and the field of
view and multiplex of the spectrograph with which they are carried
out. The numbers are in some cases merely indicative, especially for
future surveys.

\begin{table}[tb]
\begin{center}
\caption{Summary of some recent and proposed redshift surveys.}
\begin{tabular}{lrccccrr}
\hline
Survey & $N_{gal}$ & Band & Mag & $\overline{z}$ & Diam. & FoV & Multi \\
 & & & lim & & (m) & ($\Box^\circ$) & -plex \\
\hline
CfA     &    15000 & $B$      & 15.5 & 0.02 &  1.5 &   --- &     1 \\
SSRS    &    10000 & $B$      & 15.5 & 0.02 &  1.5 &   --- &     1 \\
SAPM    &     1800 & $b_J$    & 17.2 & 0.04 &  2.3 &   --- &     1 \\
PSCz    &    15000 & $S_{60}$ &  0.6 & 0.03 &  2.1 &   --- &     1 \\
LCRS    &    26000 & $R$      & 17.7 & 0.10 &  2.5 &  1.50 &   112 \\
ESP     &     4000 & $b_J$    & 19.4 & 0.10 &  3.6 &  0.50 &    50 \\
ESOSc   &     1000 & $R$      & 20.5 & 0.30 &  3.6 &  0.70 &    50 \\
Autofib &     1400 & $b_J$    & 22.5 & 0.30 &  3.9 &  0.70 &    64 \\
CFRS    &      760 & $I$      & 22.5 & 0.56 &  3.6 &  0.02 &    70 \\
ORS     &     8500 & $B$      & 14.5 & 0.01 &  1.5 &   --- &     1 \\
MRSP    &   750000 & $b_J$    & 20.5 & 0.15 &  1.2 & 30.00 & 15000 \\
2dFb    &   250000 & $b_J$    & 19.7 & 0.10 &  3.9 &  2.10 &   400 \\
2dFf    &     6000 & $r_F$    & 21.0 & 0.30 &  3.9 &  2.10 &   400 \\
SDSS    &  1000000 & $g$      & 18.3 & 0.10 &  2.5 &  3.00 &   640 \\
FLAIR   &   100000 & $J$      & 13.0 & 0.04 &  1.2 & 30.00 &   150 \\
VLTIR   &   100000 & $K$      & 21.0 & 1.00 &  8.2 &  0.35 &   200 \\
LAMOST  & 10000000 & $B$      & 20.0 & 0.12 &  4.0 & 21.20 &  5000 \\
\hline
\end{tabular}
\end{center}
{\footnotesize {\sc Notes} (undated references are to these proceedings):
CfA = Center for Astrophysics redshift survey (Kurtz, da Costa); SSRS
= Southern Sky Redshift Survey (da Costa); SAPM = Stromlo--APM
redshift survey (Loveday \etal\ 1992); PSCz = IRAS Point Source
Catalogue $z$-survey (Rowan-Robinson); LCRS = Las Campanas Redshift
Survey (Shectman \etal\ 1996, preprint); ESP = ESO Slice Project
(Vettolani, Zucca); ESOSc = ESO-Sculptor faint galaxy redshift survey
(Bellanger \etal\ 1995); Autofib = Autofib redshift survey (Ellis
\etal\ 1996, this paper); CFRS = Canada France Redshift Survey (Lilly
\etal\ 1995, Hammer); ORS = Optical Redshift Survey (Santiago \etal\
1995); MRSP = M\"{u}nster Redshift Survey Project (Schuecker);
2dFb/2dFf = 2dF bright and faint redshift surveys (this paper); SDSS =
Sloan Digital Sky Survey (Kron); FLAIR = FLAIR southern sky survey
(Parker \etal); VLTIR = VLT near-IR redshift survey (this paper, Le
Fevre); LAMOST = Large Area Multi-Object Spectroscopy Telescope (Chu).}
\end{table}

Various figures-of-merit can be envisaged for wide-field, multi-object
spectrographs. The usual figure-of-merit for {\em imaging} surveys is
$A\Omega$: the product of the telescope aperture $A$ and the imaging
field-of-view $\Omega$. The analogous figure for spectroscopic surveys
is $AM$: the product of the telescope aperture and the multiplex $M$
of the spectrograph (i.e.\ the number of objects which can be observed
simultaneously). Note in the following exercise we are not comparing
the redshift surveys with each other (in which case a more appropriate
figure-of-merit might be the effective volume surveyed), but rather
the potential of a given telescope/spectrograph combination for
carrying out such surveys.

\begin{figure}
\epsfxsize=\textwidth
\epsfbox{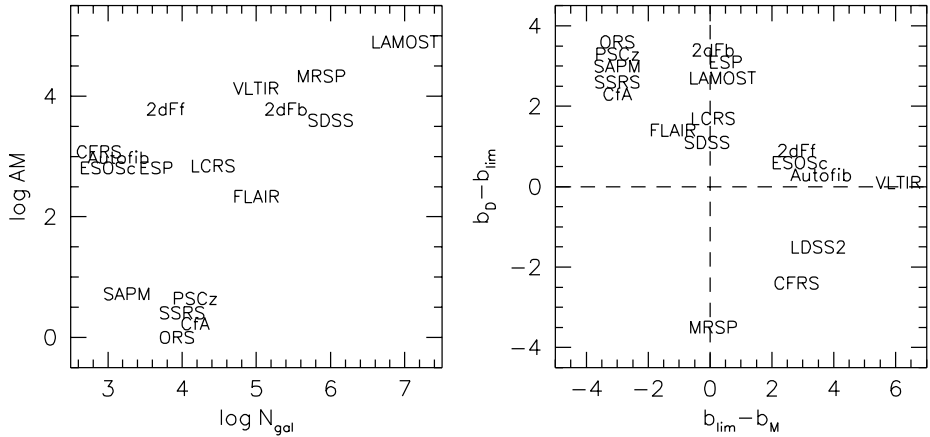}
\caption{(a)~The product of telescope aperture, $A$, and spectrograph
multiplex, $M$, versus the survey size, $N_{gal}$, for various current
and future galaxy redshift surveys. (b)~The distribution of survey
magnitude limits, $b_{lim}$, compared to the limiting magnitude of the
telescope for spectroscopy, $b_D$, and the magnitude where the number
of galaxies in the field equals the spectrograph multiplex, $b_M$.}
\end{figure}

Figure~1a shows $AM$ as a function of sample size for the surveys
listed in Table~1. The one-by-one surveys (Stromlo-APM, CfA, SSRS,
ORS, PSCz) all have log$AM$$<$1 since they were all carried out on
small (1m--2m) telescopes, yet they have sample sizes comparable to
any of the recent group of redshift surveys carried out on larger
(2m--4m) telescopes with multi-object spectrographs. This reflects
partly the large amount of time spent by dedicated observers, but,
more importantly, the small field sizes available on larger
telescopes.  This constraint has meant that (with the exception of
LCRS) most recent multi-object redshift surveys have concentrated on
going deep rather than covering a large volume. As the figure shows,
this situation is expected to change in the near future, as large
surveys are begun with spectrograph/telescope combinations with larger
$AM$ (FLAIR on the 1.2m UKST with 92 fibres and 6$^\circ$ FoV; SDSS on
a 2.5m with 640 fibres and 3$^\circ$ FoV; and the 2dF redshift survey
on a 4m with 400 fibres and 2$^\circ$ FoV). These instruments will
increase sample sizes by an order of magnitude for both bright surveys
(over 10$^5$ galaxies with $B$=17--19.5 for FLAIR, SDSS and the 2dF
bright survey) and faint surveys ($\sim$10$^4$ galaxies with $R$=21
for the 2dF faint survey). In the more distant future something like
the LAMOST concept (Chu, these proceedings) will achieve another order
of magnitude increase in $AM$, while multi-object spectrographs on
8m-class telescopes will perform surveys of 10$^5$ galaxies at yet
fainter limits.

As well as $AM$, there are a number of other secondary indicators of
capability. One of these is the range of limiting magnitudes, $b_M \ls
b_{lim} \ls b_D$, over which a particular telescope/spectrograph is an
effective survey tool. Here $b_M$ is defined by
$N$($<$$b_M$)=$M/\Omega$ (i.e.\ the magnitude at which the instrument
is well-matched to the surface density of galaxies), while $b_D
\approx 23 + 5\log(D/4)$ is the approximate limiting magnitude
achieved in one half-night on a telescope of diameter $D$ (a
psychological rather than physical limit, but an effective one
nonetheless). Figure~1b shows how the magnitude limits are chosen
within this range. `Big' surveys, for which the prime goal is sample
{\em size}, have $b_{lim} \approx b_M$, while `deep' surveys, for
which the prime goal is sample {\em depth}, have $b_{lim} \approx
b_D$. The single-object surveys, for which $b_{lim}$$-$$b_M$ is
arbitrarily set to $-3$, still have $b_D$$-$$b_{lim}$ comparable to
other `big' surveys. Two ultra-deep surveys (LDSS2 and CFRS) have
$b_D \ll b_{lim}$, as does the MRSP objective prism survey (where
extra depth is obtained at the cost of resolution).

The other indicators of capability are the accessible wavelength
range, the spectral resolution ($R \equiv \lambda/\Delta\lambda$), and
the instrumental throughput (which determines the $S/N$ achieved in
fixed time).  For redshift surveys we want the largest possible
spectral range but modest $R$ ($\ls$500 for evolution, $\ls$1000 for
structure) and S/N ($\ls$10).

\section{Redshift Surveys for Galaxy Evolution}

For redshift surveys investigating galaxy evolution, the ultimate aims
are to uncover the important physical processes in forming galaxies,
what makes galaxies biased tracers of the mass, and what part
environment plays in galaxy evolution. The more immediate aims are to
fully characterise the distributions of galaxy properties (luminosity,
mass, star-formation rate, type, etc.), to determine how these
properties depend on environment (i.e.\ local structure and
over-density), and to recover the evolution of the galaxy population
with redshift.

Results come from two types of survey:
\begin{enumerate}
\item
$b_{lim} \sim b_M$ surveys of $\sim$10$^4$ galaxies at $z$$\sim$0
(e.g.\ LCRS and PSCz) with the goal of definitively establishing the
properties of the local galaxy population.  The existing surveys are,
however, too small to fully explore type/environment dependencies;
proposed surveys of 10$^5$--10$^6$ galaxies (e.g.\ 2dFb and SDSS) will
overcome this limitation. For such surveys good photometry and uniform
selection criteria are crucial.
\item
$b_{lim} \sim b_D$ surveys of $\sim$10$^3$ galaxies at
$z$$\approx$0.5--1 (e.g.\ CFRS and Autofib) with the goal of tracking
the evolution of the galaxy population. Wide-area deep imaging is a
prerequisite for such surveys. The 2dF faint redshift survey will
cover $\sim$10$^4$ galaxies selected from deep photographic imaging,
but the next generation of surveys on 8m-class telescopes will demand
wide-area CCD imaging. In order to push beyond $z$$\sim$1 will require
a near-infrared multi-object spectrograph (see below; also Le Fevre,
these proceedings), as the familiar optical lines are redshifted
beyond 1$\mu$m and the spectral `desert' of the restframe UV is
shifted into the optical. Beyond $z$$\sim$3, optical spectroscopy is
appropriate again as Ly\,$\alpha$, CIV and the Lyman limit become
accessible.
\end{enumerate}

\section{The Autofib Redshift Survey}

The Autofib redshift survey (Ellis \etal\ 1996, Heyl \etal\ 1996),
along with the CFRS (Lilly \etal\ 1995), defines the current state of
deep field surveys. The main motivation for the survey was tracking
the evolution of the galaxy luminosity function (LF) out to
$z$$\sim$0.5 over a wide range of intrinsic luminosities. A survey
with broad coverage in both $L$ and $z$ implies the sample must span a
very broad range of apparent magnitudes. Our solution was to combine
several of our previous redshift surveys, with magnitude limits
between $B$=17 and $B$=24, with $\sim$1000 new galaxy redshifts
filling the gaps between these surveys and increasing the total sample
size.

The {\em local} LF that we obtain from the 291 galaxies with
$0.02<z<0.1$ is adequately represented by a Schechter function over
the range $-20<M<-14$. The best-fit parameters are:
$M^*$=$-$19.3+5log$h$, $\phi^*$=0.026\,$h^3$\,Mpc$^{-3}$,
$\alpha$=$-$1.1. This fit is compared in Figure~2 to similar
determinations of the local LF obtained in other studies. The
significant differences between these results are not
understood. Specific problems with some individual surveys have been
noted (e.g.\ clustering in the DARS survey, Zwicky magnitudes and the
presence of Virgo in the CfA survey), while the discrepancies between
the other surveys have variously been ascribed to magnitude scale
errors, different isophotal magnitude systems, ultra-large scale
structure, and very rapid evolution at low redshift. None of these
possibilities are very palatable, nor particularly credible. The
solution to this problem is {\em not} bigger redshift surveys (except
that these will tie down the faint end of the LF somewhat more
securely) but rather better samples, with low-isophote photometry of
uniform quality and calibration over large areas, and selection
criteria that are as inclusive as possible (particularly with respect
to low surface brightness galaxies).
 
\begin{figure}
\epsfxsize=\textwidth
\epsfbox{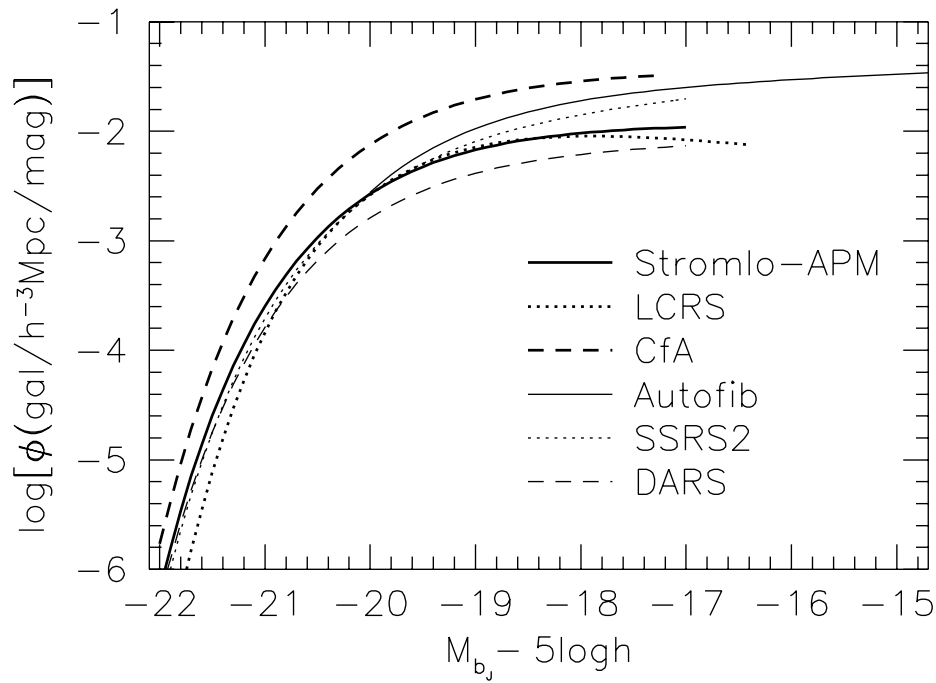}
\caption{A comparison of the Schechter function fits to various
determinations of the local galaxy luminosity function. We put all the
luminosity functions on the $b_J$ system, assuming $b$$-$$r$=1.1 for
LCRS and $b$$-$$b_{Zwicky}$=$-$0.7 for CfA (to give the same $M^*$ as
the others). See caption to Table~1 for survey references; DARS is
Efstathiou \etal\ (1988).}
\end{figure}

The broad magnitude range and faint limiting magnitude of the combined
Autofib redshift survey gives unambiguous evidence for LF evolution
with redshift. The faint end of the overall LF becomes progressively
steeper with increasing lookback time (see Figure~10 of Ellis \etal\
1996), so that by $z$$\sim$0.5 there is a noticeable change in the LF
around $L^*$. A $\chi^2$ test rejects a non-evolving LF with a
probability $<$10$^{-20}$. Although evolution is clearly significant
for $z\gs0.3$, whether there is evolution at $z\ls0.3$ remains
uncertain due to the uncertainties in the local LF. The change in the
LF shape is due to the rapid evolution of star-forming galaxies with
high [OII] equivalent widths. Figure~3 shows that the median
equivalent width of [OII] amongst late-type spiral galaxies increases
with redshift at fixed luminosity, implying higher rates of
star-formation in all objects, with bright galaxies showing this
increase at higher redshift than faint galaxies. The evolution of the
LF for each spectral type (see Heyl \etal\ 1996) reflects this effect:
the E/S0 LF shows no significant evolution out to at $z$$\sim$0.7; the
early-spiral LF shows a steepening of the faint-end slope, though the
number of $L^*$ galaxies does not change by $z$$\sim$0.5; the
late-spiral LF evolves strongly, with even $L^*$ galaxies increasing
in number by $z$$\sim$0.5. We conclude that the evolution of the
population to $z$$\sim$0.5 is completely dominated by late types.

\begin{figure}
\epsfxsize=\textwidth
\epsfbox{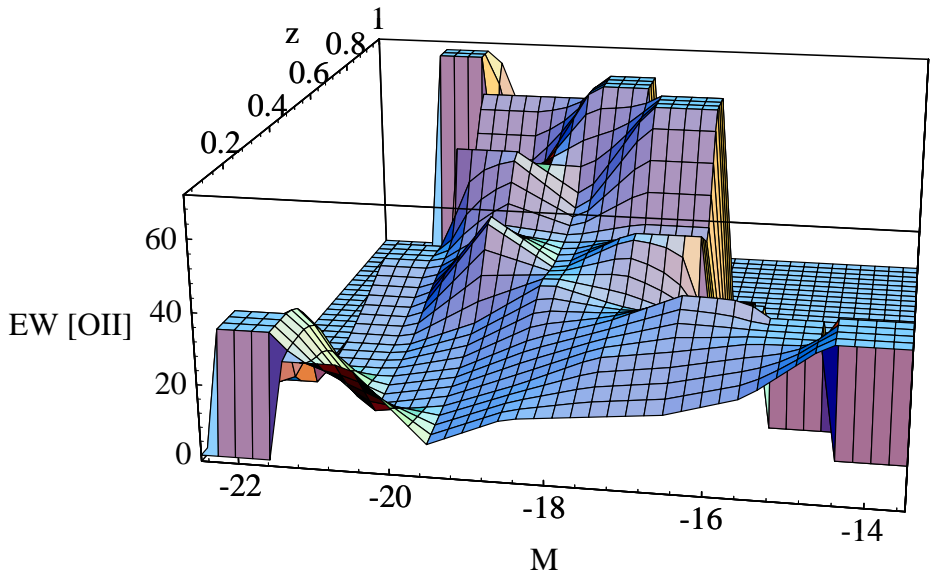}
\caption{The median [OII] equivalent width (an indicator of
star-formation rate) as function of luminosity and redshift for the
late-type spirals in the Autofib survey.}
\end{figure}

\section{Cluster Redshift Surveys}

Current redshift surveys in clusters are focussed on cluster structure
and dynamics or the evolution of galaxies in clusters. There is a
clear linkage between these topics, since the detailed dynamical
analysis of cluster substructure can reveal the formation history of
the cluster, and hence the environmental effects suffered by the
individual galaxies. Another motivation for some recent cluster
surveys (e.g.\ the CNOC survey; Carlberg \etal\ 1996) is to map the
infall pattern of galaxies around clusters, which can be used to
estimate $\Omega_{\rm cluster}$.

Again, various types of survey are appropriate depending on the
specific goals.  Complete redshift surveys of single clusters to faint
limits ($\sim$10$^3$ galaxies) can yield the detailed formation
history of the cluster and the dynamics of the galaxies as a function
of type and luminosity.  By contrast, surveys of $\sim$10$^2$ galaxies
in each of $\sim$10$^2$ clusters allow one to study cluster dynamics
and populations as function of richness/overdensity. They can also be
used (as in the CNOC survey) to average over structures and obtain
$\langle\Omega\rangle$ from infall patterns at large radii. Finally,
deep surveys of distant clusters yield the evolution of populations
and dynamics.

\section{The Merger History of the Coma Cluster}

By measuring 243 new redshifts in Coma to obtain a sample of of 465
cluster members with velocities, Colless \& Dunn (1996) were able to
recover the merger history of the cluster from kinematic evidence.
The line-of-sight velocity distribution in Coma proves to be highly
non-Gaussian, which a new test for localised velocity structure shows
is caused by the existence of a dynamically distinct group around
NGC~4839. A maximum likelihood mixture model yields a robust partition
of the sample into a main cluster around NGC~4874 and this NGC~4839
group, which accounts for about 10\% of the total mass. A linear
two-body orbit model indicates that the two clusters are at 74$^\circ$
to the line of sight with a separation of 0.8\Mpc, and are approaching
each other at 1700\kms.  Velocity maps of the cluster core show that
there is an ongoing merger between the main cluster and another group
around NGC~4889.

Several lines of evidence, including the substructure revealed in
velocity space, the distribution of X-ray gas and lack of a cooling
flow, the diffuse radio halo, and the cD envelopes around NGC~4874 and
NGC~4839 but not NGC~4889, together imply a self-consistent merger
history for Coma. In this picture, the NGC~4839 group is falling into
Coma along the Great Wall from the direction of A1367 and is just
beginning to penetrate the main cluster.  The NGC~4889 group is
partially disrupted due to its ongoing merger with the main cluster
and has ejected NGC~4889, once its dominant galaxy.  NGC~4874 is the
original dominant galaxy of the main Coma cluster, now slightly
dislodged from the bottom of the potential well by its recent
encounter with NGC~4889.

\section{Large Scale Structure Redshift Surveys}

Redshift surveys investigating large scale structure have as their
main aims the characterisation of the structure (via the power
spectrum or correlation function and measures of the Gaussianity,
topology and morphology of the galaxy distribution) and the
determination of the overall mass density $\Omega$ and the biasing
parameter for galaxies $b$.

Specific goals in deriving the power spectrum $P(k)$ (or,
equivalently, the correlation function $\xi(r)$) are to link the
small-scale, $z$$\sim$0 galaxy $P(k)$ to the much larger-scale,
high-$z$ $P(k)$ obtained from the microwave background by COBE and
other anisotropy experiments. Defining the shape of $P(k)$ gives the
initial mass fluctuation spectrum, yielding clues ot both the nature
of the dark matter and the value of $\Omega$.  Current surveys give
$P(k)$ to a precision of $\sim$20\% at 30~\Mpc, and to within a factor
of 2 at 100~\Mpc. The expected turnover in $P(k)$ at somewhat larger
scales is only hinted at. Determining $P(k)$ accurately on these
scales will require massive (10$^5$--10$^6$) galaxy surveys at
$z$$\sim$0.1, while establishing the evolution of $P(k)$ at higher
redshifts will require large (10$^4$--10$^5$) surveys of galaxies at
$z$$\sim$1 and of QSOs at $z$$\gg$1.
 
A crucial question is whether the large-scale (linear-regime) galaxy
distribution implies an underlying Gaussian distribution of mass
fluctuations, as predicted in many inflationary cosmologies. If so,
the large-scale distribution is fully described by its power
spectrum. On smaller scales the question is whether the
topology/morphology of the distribution is consistent with the
formation of structure by gravitational instability within the
framework of some model for the dark matter and the spectrum of
initial fluctuations. This requires surveys of compact,
densely-sampled volumes.
 
The mass density, $\Omega$, and bias parameter, $b$, are tightly
coupled in the galaxy distribution, and the directly measurable
quantity in most cases (e.g.\ from the redshift-space distortions in
the distribution) is $\beta=\Omega^{0.6}/b$. A sufficiently large
survey with galaxy type information would be able to use such methods
to establish the relative biasing of the different galaxy types, a
powerful constraint on theories of galaxy formation and evolution.
 
\section{The 2dF Galaxy Redshift Survey}

The main features of the 2dF spectrograph on the 3.9m Anglo-Australian
Telescope (see Figure~4) are: (i)~a 2-degree field of view with
atmospheric dispersion compensation, resulting from a new 4-component
prime focus corrector; (ii)~400 fibres of 2 arcsec diameter controlled
by a fast robot positioner with 0.2 arcsec precision; (iii)~focal
plane imaging using a CCD camera to assist in field acquisition and
guiding; (iv)~offline configuration in $<$30~min, with two focal
planes on a tumbler mechanism eliminating deadtime for exposures
longer than this; (v)~twin spectrographs each taking 200 fibres and
giving resolutions of 1.7\AA\ to 8.3\AA\ over 3500--10000\AA;
(vi)~pipeline reductions of data to return calibrated spectra to the
observers in real-time. More details are given by Cannon (these
proceedings).

\begin{figure}
\epsfxsize=\textwidth
\epsfbox{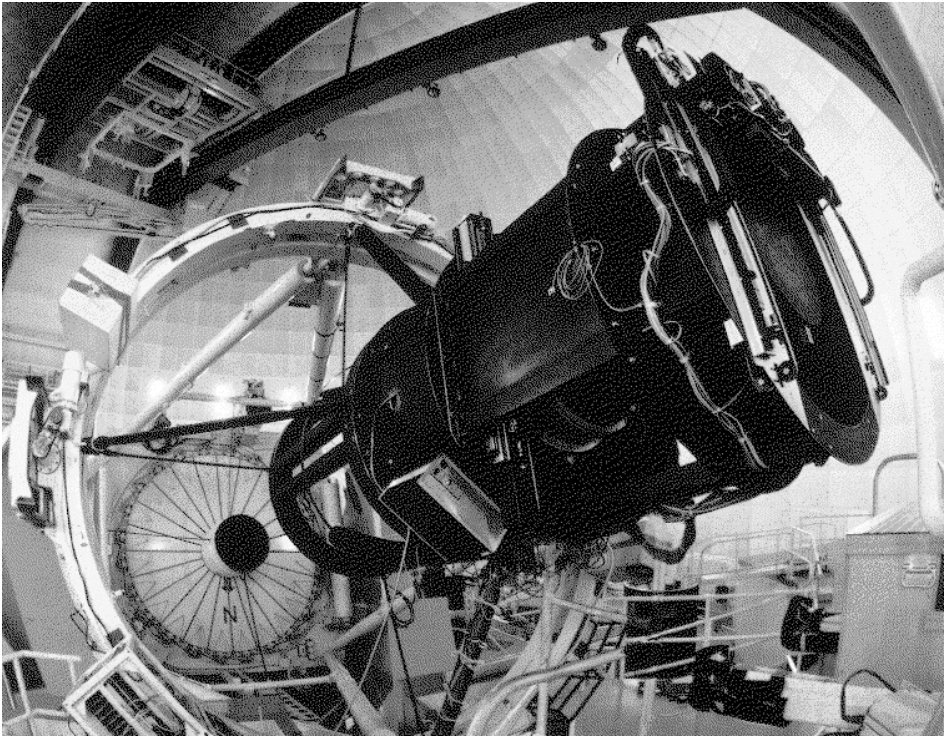}
\caption{The 2dF fibre positioner and spectrograph mounted on the
AAT. (Photo by Francisco Diego.)}
\end{figure}

Two major galaxy redshift surveys are being carried out with 2dF by a
joint Anglo-Australian collaboration: a bright ($b_J$$<$19.5) survey
of 250,000 galaxies and a faint ($R$$<$21) survey of 6000
galaxies. The overall goals of these surveys are to study large-scale
structure and galaxy evolution, and especially the interrelation of
structure and evolution. The bright survey limit of $b_J$=19.5
(extinction-corrected) implies a mean depth for the sample of
$\overline{z}$=0.1. The total area of the survey is 1700 sq.deg.,
comprising a contiguous 75$^\circ$$\times$12.5$^\circ$ strip in the
south Galactic cap, a contiguous 65$^\circ$$\times$7.5$^\circ$ strip
in the north Galactic cap, and 100 random 2dF fields over the area of
the southern APM galaxy survey.  The 250,000 galaxies will be sampled
over this area with almost 100\% completeness. The faint survey of
6000 galaxies will be carried out in override mode, taking advantage
of the highest-quality observing conditons occurring 10\% of the time
at the AAT.

We estimate that the total time required for the survey will be 90
nights. The survey will begin towards the end of 1996, and we hope to
complete it by the end of 1998.  Compared to the LCRS, the 2dF survey
has the same magnitude limit and redshift depth but 2.5$\times$ the
area, 4$\times$ the sampling rate and 10$\times$ the total number of
galaxies. The 2dF survey will be complementary to the northern SDSS
redshift survey, which has a similar magnitude limit but is planned to
cover a larger area and include up to 10$^6$ galaxies over it's
five-year nominal lifetime (Kron, these proceedings).

Future redshift surveys with 2dF will take advantage of having such a
large multiplex advantage on a 4m-class telescope to carry out massive
surveys of {\em faint} objects. For example, investing a similar
amount of time in the faint survey as the bright survey would yield
50,000 redshifts for galaxies to $R$=21, i.e.\ a survey twice as large
as LCRS but with a mean redshift of 0.3, giving a direct view of the
evolution of galaxy clustering over the last third of the history of
the universe.
 
\section{Beyond Redshift Surveys}

Redshift surveys generally use R$\ls$1000 and S/N$\ls$10 in order to
maximise the number of redshifts obtained per unit telescope
time. However, large surveys with R$\gs$3000 and S/N$\gs$20 are
needed: (i)~to measure the internal motions of galaxies to the
faintest possible limits; (ii)~to measure star-formation rates from
H$\alpha$ flux for large, representative samples; (iii)~to investigate
the star-forming dwarf population (with narrow lines) at low to
moderate redshift; (iv)~to determine variations in the Fundamental
Plane (FP) and Tully-Fisher (TF) relations as a function of
luminosity, morphological type and environment; (v)~to use the FP/TF
relations to map peculiar motions in the local universe; and (vi)~to
explore the evolution of the TF/FP relations as a constraint on galaxy
evolution.
 
Few such surveys have yet been carried out, but some first steps are
already being taken towards the latter goal. Van Dokkum \& Franx
(1996) and Bender \etal\ (1996) have used the FP in clusters out to
$z$$\sim$0.5 to constrain the evolution of elliptical galaxies and
obtain an estimate of $q_0$, while Rix \etal\ (1996) and Simard \&
Pritchet (1996) have studied the evolution of the linewidth--luminosity
relation for field galaxies in order to estimate the amount of
luminosity evolution occurring in galaxies out to $z$$\sim$0.5. At
even higher redshifts, Vogt \etal\ (1996) have used the Keck telescope
to measure rotation curves for a few galaxies out to $z$$\sim$1. This
exciting new probe of galaxy evolution is still being developed, and
larger samples are clearly needed (at low redshift as well as high
redshift) in order to fully understand the results hinted at in these
first studies.

\section{Future Prospects for Survey Spectroscopy}

There are at least three main tracks for survey spectroscopy to take:
\begin{enumerate}
\item
4m-class aperture, very wide field ($\gs$2$^\circ$), very large number
of fibres ($\gs$ several hundred): optimal for undertaking massive
($\gs$10$^6$ objects) surveys of galaxies/QSOs to B$\ls$23 and
$z$$\ls$1; this approach is currently exemplified by 2dF, with the
future perhaps looking like LAMOST (Schmidt field, 4m aperture, 5000
fibres; see Chu, these proceedings).
\item
8m-class aperture, small ($\ls$10$^\prime$) field, low multiplex
($\ls$100), `cheap' spectroscopic survey telescopes: optimal for
carrying out surveys of faint, low-density sources (e.g.\ QSOs, ALS,
radio galaxies); this approach is exemplified by the Hobby Eberley
Telescope (Sebring \etal\ 1995).
\item
8m-class aperture, moderate ($\gs$10$^\prime$) field, large ($\gs$100)
multiplex: optimal for massive surveys of very faint objects of high
surface density such as high-redshift galaxies; exemplified by the
DEIMOS optical multislit spectrograph proposed for the Keck telescope
and by the optical/near-IR multi-object spectrographs proposed for the
VLT (see below; also Le Fevre, these proceedings).
\end{enumerate}
 
\section{AUSTRALIS: a Multi-Object Spectrograph for the VLT}

AUSTRALIS is one of two concept studies for a multi-object
optical/near-infrared spectrograph optimised for high-redshift
observations on ESO's VLT (the alternate VIRMOS concept is discussed
by Le Fevre, these proceedings). A near-IR spectrograph is essential
because for $z$$>$1 most of the familiar, strong and well-studied
spectral features seen in the restframe visible have been shifted into
the near-IR. A multi-object capability is likewise essential because
sample size is critical for both population studies and structure
surveys, and also in searching for rare objects. An 8m-class telescope
is essential because normal galaxies at $z$$>$1 are very faint (the
median redshift of the galaxy population only reaches $z$=1 for
$K$=21).

Features of the AUSTRALIS design include: (i)~spectral range: 0.5--1.8
microns (V to H bands); (ii)~field of view: 20~arcmin diameter (VLT
Nasmyth); (iii)~feed mode 1: up to 400 individual 1.5~arcsec fibres;
(iv)~feed mode 2: integral field unit, FoV 9--18 arcsec, resolution
0.25--0.5 arcsec; (v)~resolving power: R$\approx$4000, permitting
digital OH sky-suppression; (vi)~detectors: six Rockwell 1024$^2$
HgCdTe devices plus two 2048$^2$ CCDs.
 
A detailed performance model suggests that feasible projects for such
a spectrograph on the VLT would include a redshift survey of 10$^4$
K$<$21 galaxies (50\% at $z$$>$1) in 10 nights and a study of the
evolution of the star-formation rate and Tully-Fisher relation via
H$\alpha$ fluxes and linewidths down to 100\kms\ for 800 galaxies over
$z$=0.6--1.7 in 4 nights. 

Such capabilities open up new horizons for studying the universe at
high redshifts, and will ensure that wide-field, multi-object
spectroscopy stays at the forefront of observational cosmology over
the next decade and beyond.
 
~\\{\bf References}\\~\\{\footnotesize
Bellanger C., de Lapparent V., Arnouts S., \etal, 1994, A\&AS, 110, 159 \\
Bender R., Ziegler B., Bruzual G., 1996, ApJ, accepted \\
Carlberg R., Yee H., Ellingson E., \etal, 1996, ApJ, 462, 32 \\
Colless M.M., Dunn A.M., 1996, ApJ, 458, 435 \\
Cowie L.L., Songaila A., Hu E.M., Cohen J.G., 1996, ApJ, submitted \\
Efstathiou G., Ellis R.S., Peterson B.A., 1988, MNRAS, 232, 431 \\
Ellis R.S., Colless M.M., Broadhurst T., Heyl J., Glazebrook K.,
1996, MNRAS, 280, 235 \\
Heyl J.S., Colless M.M., Ellis R.S., Broadhurst T.J., 1996, MNRAS,
submitted \\
Lilly S.J., Tresse L., Hammer F., Crampton D., Le Fevre O., 1995, ApJ,
455, 108 \\
Loveday J., Peterson B.A., Efstathiou G., Maddox S.J., 1992, ApJ, 390,
338  \\
Rix H.-W., Guhathakurta P., Colless M.M., Ing K., 1996, MNRAS,
submitted \\ 
Santiago B.X., Strauss M.A., Lahav O., \etal, ApJ, 446, 457 \\
Shectman S.A., Landy S.D., Oemler A., \etal, 1996, ApJ, accepted \\
Sebring T.A., Adams M., Ramsey L.W., 1995, BAAS, 187, \#121.03 \\
Simard L., Pritchet C.J., 1996, ApJL, submitted \\
Steidel C., Giavalisco M., Pettini M., Dickinson M., Adelburger
K., 1996, ApJ, 462, L17 \\
van Dokkum P.G., Franx M., 1996, MNRAS, accepted \\
Vogt N.P., Forbes D.A., Phillips A.C., \etal, 1996, ApJL, accepted \\
}

\end{document}